\documentclass{article}

\begin{document}

\title{
Gravity-induced instability of an electron on the D-3-brane}         % Enter your title between curly braces
\author{Eugen \v{S}im\'{a}nek  \footnote {electronic address: simanek@ucr.edu} \\Department of Physics, University of California, Riverside, Calif. 92521}  % Enter your name between curly braces

\date{}          % Enter your date or \today between curly braces
\maketitle

\begin{abstract}

We study the effect of gravity on stability of an electron living on the 3-brane.  Our recently proposed fermion self-energy model is augmented by the energy of self-gravitating electron.  This energy is evaluated using the gravitational field of a spherical source on the 3-brane derived by Garriga and Tanaka.  Owing to the smallness of the electron radius, the short-range correction to this field dominates the Newtonian gravitational field of the electron.  The magnitude  of the short-range correction is proportional to the square of the curvature radius, $ l $, of the anti-de Sitter space.  On increasing $l$ beyond a critical value $l_{c}$, the equilibrium radius of the electron, undergoes a transition to a region where it becomes complex-signaling an instability of the electron.  For the gravity-modified fermion self-energy model, we find $l_{c} = 2 \times 10^{-21}$ cm.  If the electron is confined by the "hedgehog" potential, the critical length is $l_{c} = 1.6 \times 10^{-12}$ cm.
\end{abstract}

\section{Introduction}

In a recent paper [1], we considered the model of an electron distributed on the surface of a cavity in the condensate of Higgs bosons.  The surface tension of this cavity prevents the electron from flying apart under the action of Coulomb repulsion.  In 1962, Dirac [2] proposed a similar model though without reference to the Higgs condensate.  The main motivation for his work was to explain the muon as the excited state of the vibration of the electron radius about its equilibrium.  By neglecting the vibrational energy, the total energy of the electron consists of the Coulomb part and the surface energy of the cavity [2].  The total energy is minimized when the equilibrium radius is of order $10^{-13}$ cm which is about $10^{5}$ times larger than recent experimental data [3].  In our paper [1], we addressed this problem by replacing the Coulomb energy by fermion self-energy involving screening by electrons occupying the negative energies of the vacuum [4].

In Ref. [1], we point out that the equilibrium radius of the electron is small enough $(R_{0} \approx 10^{-31} \textrm {cm})$ to make the gravitational forces play a significant role.

In Sec. 2 of the present paper, we calculate the gravitational energy of a spherical electron mass using the modified Newtonian potential derived by Garriga and Tanaka [5].  These authors studied the gravitational field created by isolated matter sources in the Randall-Sundrum-brane world [6, 7].  According to Ref. [5], the potential which describes attraction of neighboring bodies is a sum of the Newtonian term proportional to $1/r$ and a short-range correction which goes as $l^{2}/r^{3}$ where $l$ is the curvature radius of the AdS.  In Sec. 2, we argue that owing to the smallness of the equilibrium electron radius, the short-range correction dominates the gravitational potential.

Using this potential, we calculate the energy of the self-gravitating sphere following the method used by Landau and Lifshitz [8].

In Sec. 3, we study the effect of gravity on the equilibrium radius of an electron living on the 3-brane.  We start with the fermion self-energy model (Eq. (62) of Ref. [1]) augmented by the gravitational energy derived in Sec. 2 of the present paper.  The resulting total energy, $E_{tot}$, is used to obtain the equilibrium electron radius by setting $\partial E_{tot}/\partial R = 0$.

Sec. 4 describes the instability of electron radius induced by gravity.  We find that upon increasing the length $l$ beyond certain critical value, $l_{c}$, the equilibrium radius of the electron undergoes a transition from real to complex value.  For the fermion self-energy model, the value of $l_{c}$ is of order $10^{-21}$ cm.

In Sec. 4, we also investigate another model which allows the coherence length, $\xi_{0}$, to be comparable or even larger than $R_{0}$.  In this model, the electron is confined by the "Higgs" potential well,  similar to the "hedgehog potential" of Polyakov [9] and 't Hooft [10], proposed previously for the magnetic monopole (see also Peebles [11]).  Upon increasing $l$, the electron radius decreases toward zero at critical length $l_{c}$.  

The critical length, $l_{c}$, found from analyzing the corresponding equation for the equilibrium radius of this model, is of order $10^{-12} \textrm {cm}$.

Sec. 5 is devoted to a discussion of possible cosmological ramifications of the instability found in this work.

\section{Energy of Self-Gravitational Sphere}

In this section, we study the gravitational potential energy, $E_{gr}$, for a mass $M$ contained within a sphere of radius $R$.  Assuming a homogeneous mass distribution with density $\rho_{0} (R)$, we have 

\begin{equation}\label{Eq1}
M = \frac{4 \pi}{3} R^{3} \rho_{0}(R)
\end{equation}

Following Ref. [8], we first consider the mass $M(r)$ within a sphere of radius $r$ and obtain the work required to bring additional  mass $d M (r)$ of a spherical shell of radius $r$.  For Newtonian gravity, this work is [8]

\begin{equation}\label{Eq2}
-G_{N}M (r) \frac{dM(r)}{r}
\end{equation}

where

\begin{equation}\label{Eq3}
M (r) = \frac{4 \pi}{3} r^{3}\rho_{0}
\end{equation}

Using Eqs. (2) and (3), the potential energy $E_{gr}$ is

\begin{equation}\label{Eq4}
E_{gr} = - G_{N}\int ^{R}_{o} \frac{M(r) d M(r)}{r}  = - \frac{3 G_{N}}{5} \frac{M^{2}}{R}
\end{equation}

In the present paper, the electron is constrained to live on a domain wall (3-brane) which is embedded in a five-dimentional anti-de Sitter space (AdS)[6, 7].
Garriga and Tanaka [5] considered the gravitational field of a spherically symmetric source on the 3-brane.  According to Eq. (23) of Ref. [5], the potential which determines the attraction of neighboring bodies is

\begin{equation}\label{Eq5}
\frac{h_{00}}{2} = \frac{G_{N }M}{r} \Big(1 + \frac{2l^{2}}{3r^{2}} \Big)
\end{equation}

The correction to the Newtonian potential, given by the second term in the parenthesis, is due to the Kaluza-Klein corrections to the linearized gravitation field  of a nonrelativistic spherical object (see also Ref. [12] for another derivation of the short-range corrections to the gravitation interaction on the brane).  We note that $l$ is related to the cosmological constant in the bulk, $\Lambda$, by $\Lambda = - 6 l^{-2}$ [5].  In other words, $l$ is the curvature radius of AdS (see Sec. 4).

Eq. (5) suggests that for an electron living on the brane, the work required to add a spherical mass-shell of radius $r$ to $M(r)$ is given by

\begin{equation}\label{Eq6}
-G_{N} M (r) \frac{d M(r)}{r} \Big ( 1 + \frac{2 l^{2}}{3 r^{2}} \Big )
\end{equation}

Since the electron radius is presumably such that $r\ll l$[1], the expression (6) can be replaced by the short-range approximation

\begin{equation}\label{Eq7}
- \frac{2}{3} G_{N} l^{2} \frac{M(r) d M (r)}{r^{3}}
\end{equation}

The $r-$integration of this expression yields the gravitational potential energy, $\tilde {E}_{gr}$, for spherical mass $M$ on the 3-brane

\begin{equation}\label{Eq8}
\tilde {E}_{gr} = - \frac{2 G_{N} l^{2}}{3} \int^{R}_{0} \frac{M(r) d M(r)}{r ^{3}} dr
\end{equation}

Using Eq. (3) for $M (r)$, Eq. (8) yields

\begin{equation}\label{Eq9}
\tilde {E}_{gr} = - \frac{2}{3} G_{N} \frac{M^{2} l^{2}}{R^{3}}
\end{equation}

This expression represents the gravitational potential energy of a spherical mass of radius $R$ living on a 3-brane (see expression (70) of Ref. [1]).

\section{Equilibrium Radius of the Electron}

To study the effect of gravity on stability of an electron constrained to live on the 3-brane, we start from the fermion self-energy model described by Eq. (62) of Ref [1].  Augmenting this equation by the gravitational potential energy, given by Eq. (9), we obtain the following form for the total energy

\begin{equation}\label{Eq10}
E _{tot} = A  {\log} \big( \frac{\hbar}{mcR} \big) + \frac{4 \pi}{3} R ^{3} \Big ( \frac{\eta^{2}}{2 \xi ^{2}_{0}} \Big ) - \frac{2}{3}G_{N} \frac{m^{2} l^{2}}{R^{3}}
\end{equation}

where $A = 3 e^{2} mc/2 \pi \hbar$, and $\eta^{2} = 3 \times 10^{21}$ MeV/cm [13].  The numerical value of the coherence length, $\xi _{0}$, is not known but we assume that $\xi _{0} \ll R_{0}$, where $R_{0}$ is the equilibrium radius in the absence of the gravitational term in Eq. (10).  The equilibrium radius in the presence  of this term follows by setting $\partial E_{tot}/\partial R = 0$, yielding

\begin{equation}\label{Eq11}
 - \frac  {A}{R}+ \frac {2\pi R^{2} \eta ^{2}}{\xi ^{2}_{0}} - \frac {2 G_{N} m^{2} l ^{2}}{R^{4}} = 0
\end{equation}

Multiplying this equation by $R^{4}$ and defining  the quantity $y = R^{3}$, we obtain the following quadratic equation for $y$

\begin{equation}\label{Eq12}
y ^{2} - \frac{A \xi^{2}_{0}}{2 \pi \eta^{2}} y + \frac {G_{N} m ^{2} l ^{2} \xi^{2}_{0}} {\pi \eta^{2}} = 0 
\end{equation}

with the roots

\begin{equation}\label{Eq13}
y _{1,2} = \frac{A \xi^{2}_{0}} {4 \pi \eta^{2}} \pm \sqrt {\Big (\frac {A \xi ^{2}_{0}} {4 \pi \eta^{2}}\Big)^{2} - \frac{G_{N}m ^{2}l^{2} \xi^{2}_{0}}{\pi \eta^{2}}}
\end{equation}

When the gravitational term is neglected, the positive root of (13) yields

\begin{equation}\label{Eq14}
y_{1} = \frac{A \xi^{2}_{0}}{2\pi \eta^{2}}
\end{equation}

which implies that the equilibrium radius $R_{0}$ is given by

\begin{equation}\label{Eq15}
R^{3}_{0} = \frac{3 e^{2}\xi^{2}_{0}mc^{2}}{ 4 \pi^{2} \eta ^{2} \hbar c}
\end{equation}

This result is in agreement with Eq. (64) of Ref. [1].  Hence, we confine ourselves to the positive root of Eq. (13).  

For non-zero gravitational  term , the equilibrium radius is then given by

\begin{equation}\label{Eq16}
\Big ( \tilde {R}_{0} \Big )^{3} =  \frac {R^{3}_{0}}{2} \Bigg (  1 + \sqrt {1 - \frac {4 G_{N}m^{2}l^{2} \xi^{2}_{0}}{\pi \eta ^{2}R ^{6}_{0}}}\Bigg)
\end{equation}

\section {Gravity-induced instability of $\tilde {R}_{0}$}

From Eq. (16), we see that gravity tends to reduce $(\tilde {R})^{3}_{0}$. As long as

\begin{equation}\label{Eq17}
\frac {4 G_{N}m^{2} l^{2}\xi ^{2}} {\pi \eta^{2} R^{6}_{0}} \leq 1
\end{equation}

$(\tilde{R}_{0})^{3}$ remains real.  On the other hand, when the quantity on left-hand side of (17) is larger than 1, $(\tilde{R} _{0})^{3}$ becomes a complex number.  Solving the equality (17) for $l$, we arrive at the critical value of $l$ satisfying

\begin{equation}\label{Eq18}
l^{2}_{c}= \frac {\pi \eta^{2}R^{6}_{0}} {4 G_{N}m^{2}\xi^{2}_{0}} = \frac{10^{6}\pi \eta^{2}R^{4}_{0}} {4 G_{N} m^{2}}
\end{equation}

where the second equality follows by assuming $\xi_{0} = 10^{-3} R_{0}$ and taking $\eta^{2} = 3 \times 10 ^{21}$ MeV/cm (Ref. [13]).  For $R_{0} = 9 \times 10 ^{-32}$ cm and $m c^{2} = 0.51$ MeV, the right-hand side of Eq. (18) can be evaluated yielding

\begin{equation}\label{Eq19}
l_{c} \approx  2 \times 10^{-21} \textrm {cm}
\end{equation}

The energy corresponding to $l_{c}$ can be estimated from the relation

\begin{equation}\label{Eq20}
\epsilon_{c} = \frac{\hbar c}{l_{c}}= 10^{7} \textrm {GeV}
\end{equation}

We note that this energy is located in the "desert epoch" of the "history" of the universe [13].

It should be pointed out that the results (19) and (20) stem from the assumption, $\xi_{0} \ll R_{0}$ (made in Ref [1]).

In view of the fact that the Ginzburg-Landau coherence length $\xi_{0}$  is not known, the assumption, $\xi_{0}\ll R_{0}$, is not well founded.  For that reason, it may be instructive to investigate another model in which the confining potential for the electron stems from the "hedgehog" configuration of the Higgs field.  This leads to the following form of total energy [1, 11]

\begin{equation}\label{Eq21}
E_{tot} = A \log \frac{\hbar} {mcR} + 4 \pi \eta^{2} R - \frac{2}{3} G_{N} \frac{m^{2}l^{2}} { R^{3}}
\end{equation}

where the second term corresponds to the hedgehog mass within distance $R$ of the core [11].  The equilibrium radius of the electron, $R_{0}$, (in the absence of gravity) is determined  by setting $\partial E_{tot}/\partial R = 0$ where $E_{tot}$ is given by the first two terms of Eq. (21).  In this way, we obtain

\begin{equation}\label{Eq22}
R_{0} \simeq \frac{A}{4 \pi \eta^{2}} \simeq 4.7 \times 10 ^{-26} \textrm {cm}
\end{equation}

In the presence of gravity, the equilibrium radius is found from full equation (21), yielding

\begin{equation}\label{Eq23}
\frac{\partial E_{tot}}{\partial R} = - \frac{A}{R} + 4 \pi \eta^{2} + \frac{2 G_{N} m^{2}l^{2}}{R^{4}}
\end{equation}

Multiplying Eq. (23) by $\frac {R^{4}}{4 \pi \eta^{2}}$, we obtain

\begin{equation}\label{Eq24}
R^{4}- \frac{A}{4 \pi \eta^{2}} R^{3}+ \frac{G_{N} m^{2}l^{2}}{2 \pi \eta^{2}} = 0
\end{equation}

To make an orientational estimate of the critical length $l_{c}$, we solve Eq. (24) to first order in the gravitational perturbation $G_{N}m^{2}l^{2}/2 \pi \eta^{2}$.

Expressing the perturbed radius $R$ as

\begin{equation}\label{Eq25}
R = R_{0} (1 + x)
\end{equation}

we obtain from Eq. (24)

\begin{equation}\label{Eq26}
x  \approx - \frac{G_{N} m^{2}l^{2}} {2 \pi \eta^{2} R^{4}_{0}}
\end{equation}

From Eqs. (25) and (26), we see that $R > 0$ as long as $|x| < 1$. This gives us a rough estimate of $l_{c}$.  Using Eq. (26), we have

\begin{equation}\label{Eq27}
l_{c}= \Bigg [\frac{2 \pi \eta ^{2}R ^{4}_{0}}{G_{N}m^{2}} \Bigg ]^{\frac{1}{2}} \simeq 1.6 \times 10 ^{-12}\textrm {cm}
\end{equation}

\section {Cosmological ramifications}

In his book "Cycles of Time," Roger Penrose discusses various problems that might occur in the late stages of the evolution of universe in the scenario of cyclic cosmology.  His particular concern has to do with the decay of electrons and positrons.

Perhaps, the gravity-induced instability, discussed in Sec. 4 of the present work, could provide the desired mechanism for this decay.

We start with the fermion self-energy model for which the instability of electron radius occurs when $l^{2} = l^{2}_{c}$, where $l^{2}_{c}$ is given by Eq. (18).  Our primary question is which parameter, in this equation, is the one that triggers this instability.  First, we focus on the pameter $\eta^{2}$, which also exhibits an instability due to quantum fluctuations of the phase analogous to those of granular superconductor [15].  These fluctuations tend to decrease the order parameter $\eta$ and make it to vanish at a critical value of the ratio of the charging energy of the grain to the Josephson integrain energy [15], [16].  This prompt us to investigate the dependence of $l^{2}_{c}$ on the parameter $\eta^{2}$.  This dependence goes, according to Eq. (18), as follows

\begin{equation}\label{Eq28}
l^{2}_{c}\propto \frac{\eta^{2}R^{6}_{0}}{m^{2}}
\end{equation}

Invoking Eq. (15), and the relation $m \propto \eta$ [13], Eq. (28), shows that

\begin{equation}\label{Eq29}
l^{2}_{c}\propto \frac{1}{\eta^{2}}
\end{equation}

Since $\eta^{2} \rightarrow 0$ at the instability of spontaneous symmetry breaking [15], Eq. (29) predicts that $l^{2}_{c} \rightarrow \infty$ unless the length $l$ undergoes a rapid increase such that $l^{2}> l^{2}_{c}$.

In a similar way, we can investigate the $\eta$-dependence of Eq. (27), yielding

\begin{equation}\label{Eq30}
l^{2}_{c} \propto \frac{\eta^{2}R_{0}^{4}}{m^{2}} \propto \frac {\eta^{2}}{m^{2}} \frac{A^{4}}{\eta ^{8}}
\end{equation}

where Eq. (22) has been used on the right-hand side.  Now since $A \propto m \propto \eta$, Eq. (30) implies

\begin{equation}\label{Eq31}
l^{2}_{c}\propto \frac{1}{\eta^{4}}
\end{equation}

Again, since $\eta \rightarrow 0$ at the instability of spontaneous symmetry breaking, implying that $l^{2}_{c} \rightarrow \infty$ unless $l \rightarrow \infty$ such that $l^{2} > l^{2}_{c}$.

In conclusion, the gravity-induced instability of the electron radius described in Sec. 4 can take place only if there is an accelerated expansion of the radius of AdS.

% Set the ending of a LaTeX document

\begin{thebibliography}{19}

\bibitem{1}
Eugen~\v{S}im\'{a}nek, 
\emph{arXiv: 1502.00983 v1}, [physics.gen-ph] 29, Jan, 2015.

\bibitem{2}
P.~A.~M. Dirac,
\emph{}
Proc. Roy. Soc. \textbf{268A}, 57 (1962).

\bibitem{3}
D.~Bourilkov,
\emph{}
Phys. Rev. D \textbf{64}, 071 701 R (2001).

\bibitem{4}
V.~F.~Weisskopf,
\emph{}
Phys. Rev. \textbf{56}, 72 (1939).

\bibitem{5}
J.~Garriga and T.~Tanaka,
\emph{}
Phys. Rev. Lett. \textbf{84}, 2778 (2000).

\bibitem{6}
L.~Randall and R.~Sundrum,
\emph{}
Phys. Rev. Lett. \textbf{83}, 4690 (1999).

\bibitem{7}
L.~Randall and R.~Sundrum,
\emph{}
Phys. Rev. Lett. \textbf{83}, 3370 (1999).

\bibitem{8}
L.~D.~Landau and E.~M.~Lifshitz,
\emph{Statistical Physics}, (Pergamon Press, Oxford, 1959).

\bibitem{9}
A.~M.~Polyakov,
\emph{}
JETP Lett. \textbf{20}, 194 (1974).

\bibitem{10}
't Hooft,
\emph{}
Nucl. Phys. \textbf{B79}, 276 (1974).

\bibitem{11}
P.~J.~E. Peebles,
\emph{Principles of Physical Cosmology}, (Princeton University Press, Princeton, New Jersey, 1993).

\bibitem{12}
M.~Gasperini,
\emph{Elements of String Cosmology},(Cambridge University Press, Cambridge, 2007).

\bibitem{13}
P.~D.~B.~Collins, A.~D.~Martin and E.~J. ~Squires,
\emph{Particle Physics and Cosmology}, (John Wiley \& Sons, New York, 1989).

\bibitem{14}
Roger~Penrose,
\emph{Cycles of Time},(Alfred A. Knopf, a division of Random House, Inc., New York, 2010).


\bibitem{15}
Eugen~\v{S}im\'{a}nek,
\emph{}
Phys. Rev. \textbf{B22}, 459 (1980).

\bibitem{16}
Eugen~\v{S}im\'{a}nek, (unpublished).








\end{thebibliography}
\end{document}